# Nationwide frequency-dependent seismic site amplification models for Iceland

Atefe Darzi[1*], Benedikt Halldorsson[1,2], Fabrice Cotton[3,4], Sahar Rahpeyma[5]

**Abstract**

Seismic wave amplification due to localized site conditions or site-effects, is an important aspect of regional seismic hazard assessment. In the absence of systematic studies of frequency-dependent site-effects during strong Icelandic earthquakes, various local site proxies of large-scale studies in other seismic regions have been used and/or proposed for application in Iceland. Recently, however, earthquake site-effects were rigorously quantified for 34 strong-motion stations in Southwest Iceland for the first time and correlated to distinct Icelandic geological units of hard rock, rock, lava rock, and sedimentary soil. These units are prevalent throughout Iceland and in this study therefore, we present 1) nationwide maps of proxies (e.g. slope, Vs30, geological units) that may contribute to a better estimation of site effects and associated, 2) frequency-dependent site-amplification maps of Iceland relative to the median ground-motion prediction of Rahpeyma et al. (2023). We particularly focus on the two transform zones of the country, on the basis of digital elevation models and detailed geological maps. Specifically, the frequency-dependent site factors for each geological unit are presented at 1, 2, 5, 7, 10-30 Hz, and PGA. Finally, for comparison, we generate site amplification maps based on recent large-scale models developed in other seismic regions and/or applied in other studies (e.g., ESRM20) as well as various site proxies they are based on (e.g., geology- and slope-based inferred $V_{S30}$, geomorphological sedimentary thickness). We compare site-proxy maps and amplification maps from both Icelandic and large-scale, non-Icelandic, models. Specifically, neither spatial patterns nor amplification levels in either proxy or amplification maps from large-scale non-Icelandic studies resemble those observed from local quantitative strong-motion research as presented in this study. We attribute the discrepancy primarily to the young geology of Iceland and its formation history. Additionally, this study compares model performance across frequencies by assessing the bias of model predictions against empirical site amplifications in the South Iceland Seismic Zone, accounting for site-to-site variability of residuals indicating the superior performance of the local amplification model. The results presented in this study now allow a more informed estimation of earthquake ground motion amplitudes on various geological units that are expected to result in more reliable and information-based seismic hazard estimates in Iceland.

[1] Faculty of Civil and Environmental Engineering, School of Engineering and Natural Sciences, University of Iceland, Reykjavik, Iceland (A.D.: atefe@hi.is, corresponding author; B.H.: skykkur@hi.is)
[2] Volcanoes, Earthquakes & Deformation, Service and Research Division, Icelandic Meteorological Office, Reykjavík, Iceland (benedikt@vedur.is)
[3] Seismic Hazard and Risk Dynamics, Helmholtz Centre Potsdam – German Research Centre for Geosciences GFZ, Telegrafenberg, 14473 Potsdam, Germany (fcotton@gfz-potsdam.de)
[4] Institute for Geosciences, University of Potsdam, Karl-Liebknecht-Str. 24-25, 14476 Potsdam, Germany
[5] Faculty of Geo-Information Science and Earth Observation, Department of Applied Earth Sciences, University of Twente, Enschede, Netherland (s.rahpeyma@utwente.nl)

## 1. Introduction

Iceland is the most seismically active country in northern Europe, situated on the Mid-Atlantic Ridge, where the extensional margin of the Eurasian and North American tectonic plates interacts with the Icelandic mantle plume. Historically, the largest earthquakes in Iceland occur in the two large transform zones: the South Iceland Seismic Zone (SISZ) and the Reykjanes Peninsula Oblique Rift (RPOR) located in Southwest Iceland; as well as the offshore Tjörnes Fracture Zone (TFZ) in northern Iceland (Figure 1a) (Einarsson 2014). Therefore, these are the regions of highest seismic hazard in the country (Figure 1a), codified as having a 10% probability in 50 years that the reference peak ground acceleration (PGA) exceeds 0.5 g (the standard acceleration due to Earth's gravity, equivalent to g-force) (Standards Council of Iceland, 2010). The SISZ is collocated with a relatively densely populated agricultural region and contains all critical infrastructures and lifelines of modern society. It is also capable of producing large earthquakes such as the 1912 $M_\text{w}$7.0 earthquake in the eastern SISZ, with the most recent ones are the June 2000 $M_\text{w}$6.4 and $M_\text{w}$6.5 earthquakes and the 29 May 2008 $M_\text{w}$6.3 earthquake (see yellow stars in Figure 1a) (Einarsson 1991, 2008, 2014). The TFZ, which is a largely offshore seismic zone with a relatively small population located at long distance from the faults, is believed to experience numerous earthquakes, with magnitudes up to 7.0 (Stefansson et al. 2008).

The above significant earthquake hazard and its proximity to the densely populated urban areas of the country's modern built environment, therefore, poses a substantial seismic risk in regions such as SISZ-RPOR and TFZ. Mitigating this risk effectively in these growing urban environments necessitates accurate seismic hazard assessments (SHA) for which ground motion models (GMMs) are needed to estimate the ground motion parameter scaling with earthquake magnitude, the distance of a site away from the earthquake source (see e.g., in Douglas 2018; Baker et al. 2021, and references therein). The scaling also needs to be specified for the localized site conditions due to seismic wave amplification due to the geological structure beneath the site (see e.g., Kawase 2003, and references therein). Local site characteristics (e.g., local geological features) have been well recognized as an important factor contributing to the impact of seismic ground motion on the built environment, affecting ground motion level, frequency contents and duration. For instance, at sites with mainly loose sediments, the recorded earthquake ground motion is amplified and therefore, deteriorates ground shaking's impact. Therefore, site effect studies to capture localized site amplifications are vital (e.g., Borcherdt 1970; Aki 1993; Boore and Joyner 1997).

For site-specific analyses or localized site amplification studies on limited number of sites, geotechnical or geophysical site descriptors such as S-wave velocity profile, fundamental resonance frequency, soil layer thickness, depth to basement, measured $V_{S30}$ (i.e. the time-averaged shear-wave velocity in the top 30 meters), and horizontal to vertical spectral ratio have proven useful, either individually or in combination with two or more site predictors (Nakamura 1989; Wills and Silva 1998; Fäh et al. 2003; Bonnefoy-Claudet et al. 2006; Derras et al. 2017; Darzi et al. 2019; Zhu et al. 2020; Bergamo et al. 2021). However, in reality, obtaining such detailed geotechnical and geophysical information for large-scale or regional SHA analysis is impractical and therefore, not readily available. Instead, proxies of site properties inferred from the accessible large-scale resources (e.g., national, global) such as topographical map, geological map, and various geomorphological map, serve as representatives of direct measurements. Transitioning from direct measurements to proxies introduces new challenges and large amount of uncertainty into SHA and risk, emphasizing the importance of understanding proxies. $V_{S30}$ is a main site characterization proxy that is widely used in GMMs either through site classification based on $V_{S30}$ or direct application of $V_{S30}$ estimates (e.g., Cauzzi et al. 2015; Darzi et al. 2019b). In addition to many GMMs, almost all building design standards including the Icelandic seismic design code rely on $V_{S30}$ to incorporate site conditions in the assessment of design ground motions. For large-scale studies, $V_{S30}$ is mainly inferred from other site descriptors such as topographical slope (Wald and Allen 2007b), and recently slope and geology (e.g., Thompson et al. (2014) for California; Foster et al. (2019) for New Zealand; Panzera et al. (2021) for Switzerland, Li et al. (2021) for Texas). In addition to slope and geology, terrain-based landform classification scheme based on multiple geomorphological quantities like local convexity and surface texture derived from DEM (Iwahashi and Pike 2007) was applied to characterize $V_{S30}$. For instance, Yong et al. (2012) for California, Stewart et al. (2014) in Greece, Di Capua et al. (2016) for Italy. We should note that many studies criticize the usefulness of $V_{S30}$ values as a reliable site amplification descriptor (e.g., at low frequencies, in deep sedimentary sites, regions with complex geological settings like Italy as discussed in Park and Hashash (2002), Pitilakis et al. (2006), Mucciarelli and Gallipoli (2006), and Gallipoli and Mucciarelli (2009)). It should be highlighted that $V_{S30}$ as a site proxy serves to predict ground motion at surface for the goal of seismic risk assessment. Therefore, the use of inferred $V_{S30}$, acting as a proxy of a proxy for local site amplification, introduces significant uncertainty that should be carefully considered in GMMs and SHA (Weatherill et al. 2023). In the most recent effort in the European Seismic Risk Model (ESRM20) project, three main site proxies- topographic slope, inferred $V_{S30}$ (from slope) and geological unit/era have been employed (Weatherill et al. 2023). These proxies have been integrated into SHA and risk evaluation, accounting for their respective uncertainties. In addition to traditional inferred site proxies like $V_{S30}$, recent studies are emphasizing on introducing alternative site proxies for enhancing site amplification modelling, such as Geomorphological Sediment Thickness (GST) proposed by Loviknes et al. (2024). Some studies have suggested that a site amplification model derived from the combination of geology and slope yields the most promising results, showcasing the significance of incorporating geological features in site amplification modelling (e.g., Weatherill et al. (2020b) for Japan and Loviknes et al. (2024) for Europe). Furthermore, the GST proxy has demonstrated slightly superior performance compared to slope and inferred $V_{S30}$ in capturing site amplification (Loviknes et al. 2024). While the use of inferred site proxies is advisable for regional seismic hazard studies covering extensive areas or when detailed site information is unavailable, it is necessary to conduct thorough investigations to critically evaluate their reliability and applicability in nationwide scale studies, particularly within the specific geological context of Iceland.

Iceland exhibits intricate and diverse types of young geological conditions, including various soil layers with biogenic sediments, glacial deposits, till, pumice, volcanic ash, alluvial deposits, delta bedding, sedimentary rock, lava, and old bedrock (Einarsson and Douglas 1994; Atakan et al. 1997; Bessason and Kaynia 2002). However, no measurement of $V_{S30}$ or velocity profile or any other site predictors in particular regarding the surficial geological condition are available for the purpose of quantifying site effects. For these reasons, almost all Icelandic GMMs employed in SHA effectively neglect the detailed aspects of site effects (Ólafsson and Sigbjörnsson 1999, 2004; Rupakhety and Sigbjörnsson 2009). A few that incorporate site effects do so in a largely simplified manner as binary variables of "rock" or "stiff soil" without detailed assessment (e.g., Kowsari et al. 2020). This is mostly based on qualitative assumptions or surface geology that categorizes rock sites as soil type A ($V_{S30}$ >800 m/s) vs. stiff-soil sites as soil type B (360< $V_{S30}$ <800 m/s) as per Eurocode 8 provisions (Sigbjörnsson et al. 2014). This is in stark contrast to standard international practice in which GMMs consider one or more proxies for seismic wave amplification due to localized site effects (Derras et al. 2017; Douglas 2018). Moreover, other GMMs from seismic regions worldwide applied to SHA in Iceland have all shown to be biased against the Icelandic strong motion dataset, and use site proxies that are either unknown for Iceland or incompatible with its geological and topographic characteristics (Kowsari et al. 2020).

Fortunately, recent efforts to quantify site effects have illuminated the issue of site amplification in Iceland. Studies using data from small urban strong-motion arrays, the ICEARRAY I in SISZ and ICEARRAY II in TFZ, exhibited considerable and variable site effects that dominate the spatial variation of various ground-motion intensity measure amplitudes (PGA, PSAs and energy-based parameters, e.g., arias intensity and cumulative absolute velocity) observed over small areas of both arrays that are resulted from respective geological structure (Rahpeyma et al. 2016, 2018, 2019, 2022a, b; Darzi et al. 2022). To elaborate, ICEARRAY I stations are considered uniformly rock sites with relatively flat topography, whereas ICEARRAY II, characterized by more diverse geology and topography, demonstrated a wider variation in station terms for PGA. In addition, PSA predictions show a significant increase in posterior of station terms across all periods with large inter-station standard deviation, mainly due to variable lava thickness. This emphasizes the importance of quantitatively estimating site-to-site amplification terms and their associated uncertainty in order to comprehensively understand the influence of complex geological features on ground shakings in Iceland (Rahpeyma et al. 2018, 2019).

Site effects observed at stations recording numerous earthquakes have demonstrated distinct and site-specific amplification responses. Recently, there has been an advanced assessment of site-to-site amplification term of GMMs derived from the site-to-site (inter-station) ground motion residuals, also known as site-specific station term, denoted as $\delta S2S_s$. The station term $\delta S2S_s$ captures the local frequency-dependent amplification or de-amplification effects on seismic ground motions at a given station (Kotha et al. 2016, 2018, 2020). While they are commonly utilized to improve site-specific predictions, it's essential to note that they are strictly valid only for the recording station sites (Rodriguez-Marek et al. 2013; Kotha et al. 2018). In Iceland, most recently, Rahpeyma et al. (2023) quantitatively estimated site-to-site amplification terms for PGA and 5%-damped PSA at periods of $T = 0.01$-$3.0$ s at 34 strong motion stations across SISZ by developing an empirical GMM using a multi-level Bayesian Hierarchical Model (BHM). The BHM model enables decomposing the ground motions into event, station, and event-station terms, therefore, determining the total variability that are divided into inter-event and intra-event variabilities. The intra-event is further divided into inter-station, event-station, and unexplained variability. They identified four groups of station terms by inspecting the systematic behavior of frequency-dependent site amplification at recording stations (station-to-station ground motion residuals, $\delta S2S_s$), namely: hard rock (HR), rock (R), (Holocene) Lava rock (L), and Soil (S) site groups. Observing a strong qualitative correlation between station term groups and their associated geological units, they proposed the first frequency-dependent average site amplification functions associated with key geological units in Iceland based on age and formation.

In summary, previous studies on site effects in Iceland via comprehensive statistical and physical modeling have shown that "rock" sites in Iceland can exhibit significant and frequency-dependent site effects that can vary greatly over short distances, leading to strong amplification at key resonant frequencies (Bessason and Kaynia 2002; Rahpeyma et al. 2016, 2019). However, the current seismic design criteria in the Icelandic building code do not differentiate these distinctive site conditions from the generic "rock" sites. Therefore, engineers would struggle to quantifiably assess site effects for the site in question.

Therefore, a more accurate and evidence-based incorporation of site effects across Iceland into seismic design practice requires a comprehensive site amplification model that spans the entire country. In this study, we build upon the abovementioned site effect studies in Iceland and establish frequency-dependent site amplification models for key geological units of Iceland. We compare these models with recent European and regional frequency-dependent site amplification models for Iceland using other models based on various site proxies. We examine models derived not only from conventional site predictors such as topographical slope, inferred $V_{S30}$ based on slope, and models obtained from multiple proxies such as slope and geology, but also those derived from emerging site proxies such as global geomorphological sedimentary thickness. Furthermore, in this study, we argue if the commonly used inferred site proxies such as $V_{S30}$ are reliable site descriptors for Iceland. In this pursuit, we compare the topographical- and geological-based $V_{S30}$ maps with the local estimates of $V_{S30}$ values closely correlated with the detailed geological attributes of the region. This map aims to provide a more accurate representation of site conditions and challenge the prevailing assumptions about slope-based inferred $V_{S30}$ maps in the Icelandic seismic landscape. The proposed seismic site amplification maps along with the local $V_{S30}$ map will serve as valuable tools for seismic hazard assessment and seismic design of structures in a young volcanic region of Iceland.

## 2. Key Icelandic site characteristics

### 2.1. Topography and slope

As site effects are driven by shallow geology and topography conditions, Digital Elevation Models (DEMs) and geological maps are fundamental data sets that have been widely used to assess site response mapping, in particular at the regional scale. Figure 1 shows Iceland's topography along with the distribution of Icelandic seismic and strong-motion networks, focusing in particular on the TFZ in north (Figure 1b) and SISZ-RPOR in SW Iceland (Figure 1d). The DEM topographical map of Iceland with a 20m-by-20m (2/3 arcsec) resolution has been used (Landmælingar Íslands 2021). Topographical slope calculated from DEM (Figure 1c) is another site characteristic used to determine site amplification either directly (Weatherill et al. 2023) or by estimating other proxies such as $V_{S30}$ whose measurement is time consuming and costly (Wald and Allen 2007a)).

The national seismic network, SIL, (gray triangles) of the Icelandic Meteorological Office (Stefánsson et al. 1993) and the Icelandic strong-motion network, ISMN (red triangles), operated by the Earthquake Engineering Research Centre at the University of Iceland (Sigbjörnsson et al. 2014) are shown in Figure 1(b, d). The two networks, the ISMN and SIL, have been deployed in an almost spatially exclusive pattern, in particular due to their respective focus of mapping Iceland's microseismicity and improving seismic risk management. The third type ICEARRAY are small-aperture urban strong-motion arrays that have been deployed since 2007 for the purpose of mapping spatial distribution of ground motion amplitudes and site effect variations over relatively short distances (Halldorsson et al. 2009, 2012; Rahpeyma et al. 2019; Blanck et al. 2022).

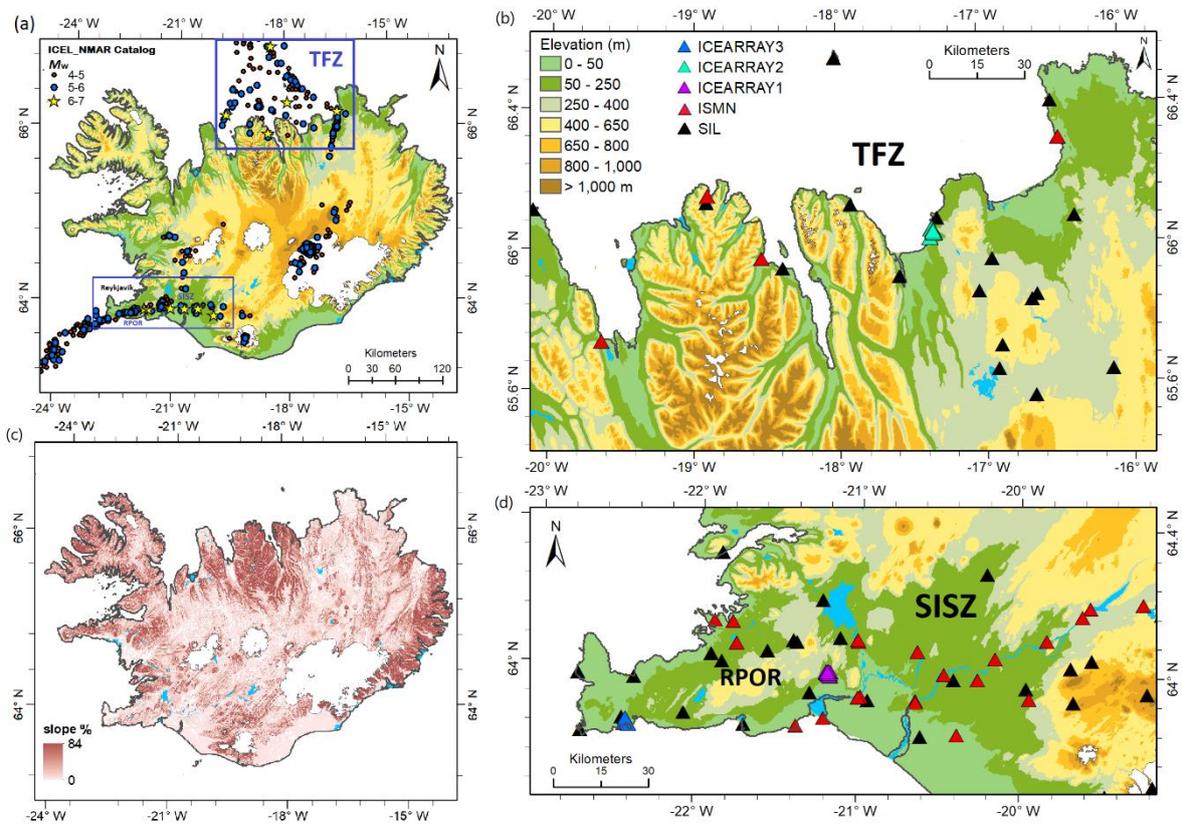

*Figure 1. (a) A map of Iceland's seismicity and topography. The two major transform zones are Tjörnes Fracture Zone (TFZ) in north Iceland and the South Iceland Seismic Zone (SISZ) and Reykjanes Peninsula Oblique Rift (RPOR) in Southwest Iceland. The circles show the spatial distribution of earthquakes with $M_w$3.3–7.1 from 1901 to 2019 (Jónasson et al. 2021). The capital city of Reykjavik, a home to two third of the country's population, is indicated on the map. (b,d) Spatial distribution of recording stations across (b) TFZ and (d) SISZ-RPOR over Iceland's topography. The locations of strong-motion recording stations of ISMN along with three strong motion arrays (ICEARRAY II in north and ICEARRAY I and III in southwest) and national seismic network (SIL) is displayed. (c) Slope map of Iceland. (For interpretation of the references to colour in this figure legend, the reader is referred to the online version of this article.)*

## 2.2. *Geology*

Besides the topography map, the geological maps of Iceland developed by the Icelandic Institute of Natural History, providing detailed descriptions of geological conditions across Iceland, offer fundamental insights into the complexities of the region's site effect study. Figure 2a shows the map of the geological age of the underlying bedrock in Iceland where strata are classified only by age, not by rock type or composition (Jóhannesson and Sæmundsson 2009). It is clear from the outset that the youngest part of the country coincides with the locations of the volcanic zones and the oldest parts are at both ends of the country in the direction of the tectonic extension. Notably, the location of the SISZ-RPOR transform zone traverses two geological age groups, with RPOR being characterized both by strike-slip faulting and extensional (volcanic) faulting (Steigerwald et al. 2020; Sæmundsson et al. 2020), the latter feature being relatively a typical of a transform fault system that develops between two offset spreading centers (Sigmundsson et al. 2020). As a consequence, the surficial geological structure of Iceland follows the geological map closely.

Figure 2c is the geological map of Iceland which broadly outlines the main features of the bedrock geology at a scale of 1:600,000 (Jóhannesson 2014) again focusing primarily on the transform zones (Figure 2b and d). this map classifies bedrock on the basis of its age, rock type, and composition. A distinction is made between volcanic and plutonic rock, and between silicic rocks (e.g., rhyolite and dacite) and basaltic and intermediate rocks (e.g., basalt and tholeiite) (Jóhannesson 2014). From older to younger, the dark blue layer, shows the oldest and hardest rock layer is basaltic rock with the age of more than 3.3 million years. The old basaltic and intermediate extrusive rocks and sediments formations from the lower Pleistocene period (0.8-3.3 million years) are indicated by a green layer. The brown layer shows the basic and intermediate hyaloclastite and pillow lava formation has been compacted since the upper Pleistocene period (younger than 0.8 million years). The acid extrusive formation created in Tertiary and Pleistocene with the age of more than 11000 years is shown with yellow color.

Due to the activity in the volcanic zones and basaltic Holocene lavas of low viscosity that were capable of flowing long distances from the points of effusion, various younger formations overlaid the older basaltic rocks. The basaltic and basic and intermediate interglacial and supraglacial lavas with intercalated sediments with age of younger than 0.8 million years (upper Pleistocene) are presented by a gray layer. Postglacial lava fields are divided into prehistoric and historic lavas (younger than AD 871). The surficial geological condition of basic and intermediate basaltic lavas, postglacial, prehistoric, older than 1100 years (> 1.1 kyr) as well as the basic and intermediate basaltic lavas, postglacial, historic with age of younger than 1100 years are presented by pink and purple layers, respectively.

It should be highlighted that Holocene lavas were considered firm rock, but over glacial-interglacial cycles, repeated sedimentary depositions and effusive eruptions, along with glacial isostatic adjustment since the last glacial period (~10 ka), have formed a complex geological structure exposing sedimentary layers beneath the lava, that could also be repeated in depth. Therefore, near the surface, high shear wave velocity resembles a hard rock site, but at depth and due to the presence of underlying softer sedimentary layers a velocity reversal is formed, sometimes intercalated due to repeating periods of deposition and volcanism.

Finally, the Holocene sediment geological condition exhibiting loose surficial sediments is depicted by a light-blue layer in Figure 2c, most notably located in the glacial runoff plains in the southeastern and southern parts of Iceland, but also in all coastal valleys albeit of lesser spatial extent.

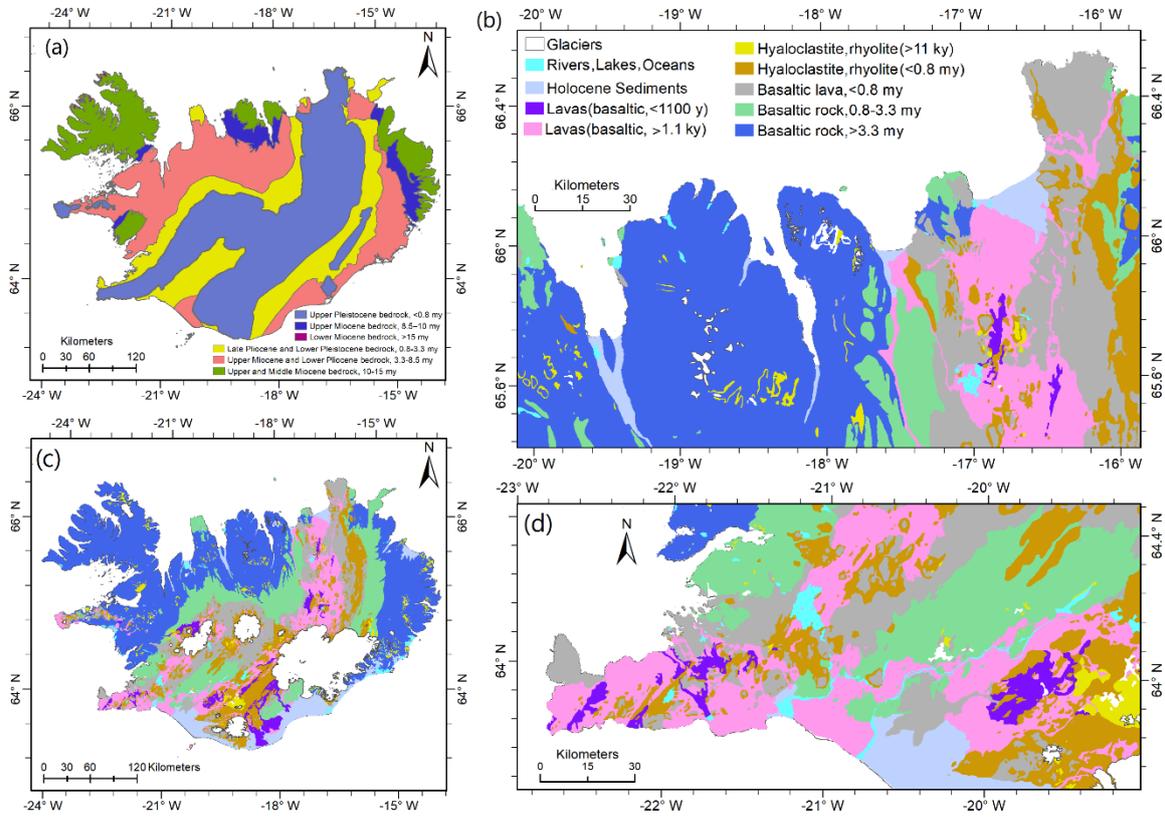

*Figure 2. (a) Map of geological age of the bedrock. (c) geological map of Iceland with magnified areas of the (b) TFZ in north Iceland and (d) SISZ-RPOR in SW-Iceland (see Figure 1). (For interpretation of the references to colour in this figure legend, the reader is referred to the online version of this article.)*

## 2.3. The $V_{S30}$ map of Iceland based on local studies

Due to the popularity that the $V_{S30}$ site-proxy has gained in the past couple of decades, and for completeness in this study, we revive and regenerate the only known $V_{S30}$ map for Iceland that is based on local studies. The work was accomplished at the Icelandic Meteorological Office during the SAFER (Seismic eArly warning For EuRope) project (Deliverable 4.36 in Vogfjörd et al. (2010), but not published in a scientific journal) and constitutes the first evidence-based $V_{S30}$ map of Iceland generated on the basis of Icelandic geological research (Figure 3). To translate the geological map into a near-surface S-wave velocity map, the generic Vs/Vp relationship of Chandler et al. (2005) was used together with available information on near-surface P-wave velocities in Iceland, obtained from measurements in boreholes, in laboratory rock samples and from refraction profiles (Gunnarsson et al. 2005). The map lists nine distinct geological units and provides their associated velocity estimates as indicated in the legend. Velocities of underwater geological units are estimated.

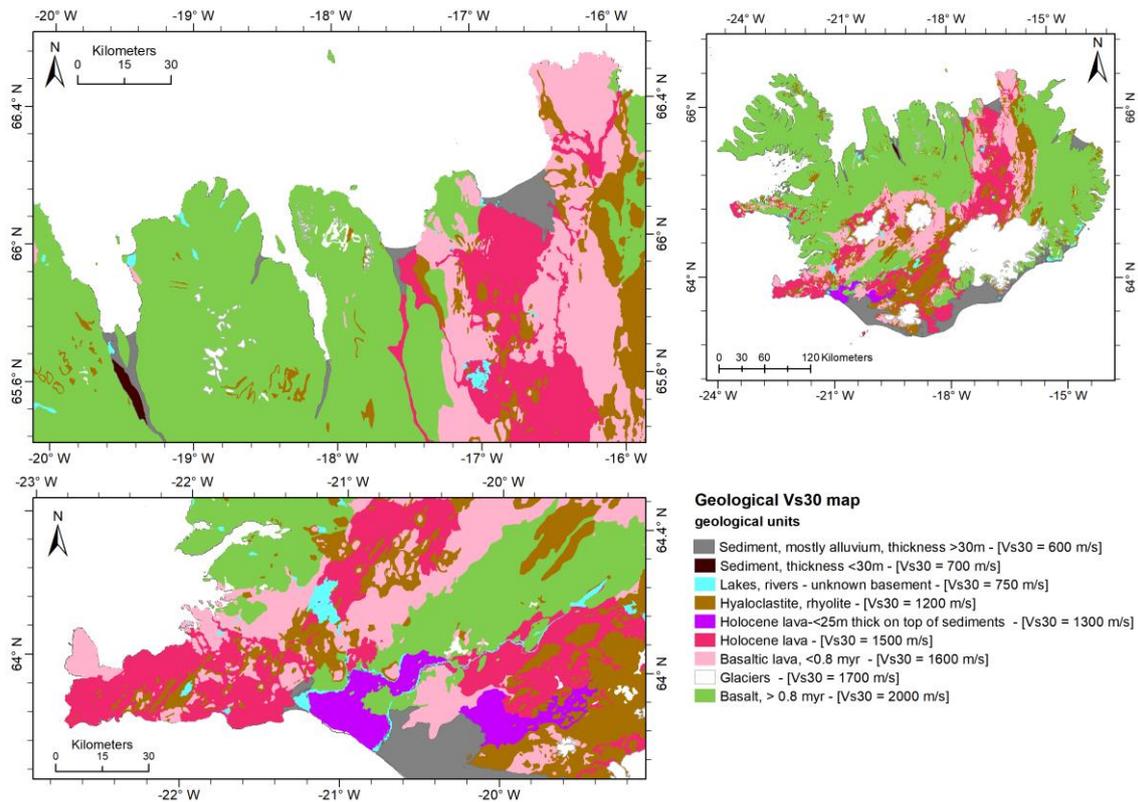

*Figure 3. Geological-based $V_{S30}$ map of Iceland (regenerated from Vogfjörd et al. (2010)). (For interpretation of the references to colour in this figure legend, the reader is referred to the online version of this article.)*

The velocities are notably higher than those of the California site-category map proposed by Wills et al. (2000), and also the near-surface velocities determined by Chandler et al. (2005). The difference is primarily attributed to the rarity of sedimentary layers, both in occurrence and in thickness, in Iceland (see the Holocene Sediments layer) and substantially metamorphosed rocks are uncommon in proximity to the surface. The older Tertiary rocks (depicted as the green layer with $V_{S30}$ ~2000 m/s) are mostly glacially eroded lava flows and the younger formations in the rift zones (illustrated as the brown and violet layers with $V_{S30}$ ~1200-1300 m/s) usually consist of fresh volcanic materials with relatively less consolidated composition. While we recognize the importance of addressing uncertainty in $V_{S30}$ estimates, it was not feasible to do so in this study as the SAFER project's estimation of the $V_{S30}$ for Iceland did not include any discussion or presentation of uncertainty associated with the P or S wave velocity estimates, nor did it provide details on the Vs/Vp relationship utilized.

## *2.4. Site proxy maps from large-scale (non-Icelandic specific) studies*

### *2.4.1. Slope-based $V_{S30}$ maps*

In large-scale site studies, regional mappable site proxies such as topographical slope, geology, … are used either to derive the site amplification predictive models or to estimate other proxies such as $V_{S30}$. One of the widely used methods to infer $V_{S30}$ is using a topographical slope. Correlating topography with measured $V_{S30}$ based on slope calculations on SRTM30 (a 30 arc-seconds land topography dataset), a first-order estimation of $V_{S30}$-based site classes (e.g., NEHRP) is obtained (Wald and Allen (2007)). The united states geological survey (USGS) developed separate models for active and stable tectonic regions, and developed a global map of inferred $V_{S30}$ ranges (Wald and Allen 2007) that is arguably the most widely used inferred site proxy to represent site amplification in GMMs and seismic hazard studies (Silva et al. 2020). Herein, we sliced the USGS global $V_{S30}$ dataset for Iceland in Figure 4a.

Lemoine et al. (2012) observed that slopes derived from land terrain models only (e.g., SRTM30 land DEM) are associated with artefacts in coastal areas resulting in several outliers (if slope is not calculated using bathymetric data). To overcome this drawback, Weatherill et al. (2023) used joint topo/bathymetric DEM of GEBCO_2014 (General Bathymetric Chart of the Oceans grid) with 30 arc-second resolution that is a global terrain model for ocean and land, and calculated the slope to provide a map of $V_{S30}$ ranges for the whole of Europe (Figure 4b). The map is utilized in the site response modelling for the latest European Seismic Risk Model, ESRM20 (Weatherill et al. 2023).

A comparison between local $V_{S30}$ values associated with Iceland's geological units (Figure 3) and the slope-based $V_{S30}$ maps of USGS and ESRM20 (Figure 4a, b) reveals significant discrepancies across the whole country, primarily due to their correlation with slope. Furthermore, the ESRM20 inferred $V_{S30}$ map was developed using a limited number of measured $V_{S30}$ values obtained from rock sites ($V_{S30}$>800 m/s), while the local $V_{S30}$ map reveals that the majority of Iceland is characterized by $V_{S30}$ values exceeding 1200 m/s.

Despite previous studies emphasizing the broad-scale applicability of the slope-based $V_{S30}$ estimation, it falls short in the nationwide context of Iceland. This failure is attributed to the necessity for detailed consideration of the geological conditions (e.g., Wald and Allen 2007; Allen and Wald 2009; Lemoine et al. 2012). To elaborate, we stress that estimating $V_{S30}$ from slope is based on the assumption that steeper slopes are predominantly formed by stiffer materials than soft soil and thus, resulting in a higher shear wave velocity. This is while on flat areas with overall low slope, e.g., basin regions, soft sediments with lower shear wave velocity are likely. This hypothesis is not valid for nominally flat-lying volcanic plateaus, carbonate rocks, and continental glacial terrains where topography alone cannot differentiate between topographically similar depositional (glacial till) drumlins and erosional (bedrock) sheepback formations (Wald and Allen 2007a). Also, previous studies show that, contrary to shallow crustal regions, the slope-based $V_{S30}$ estimates are not suitable for stable continental regions, especially in small basins, narrow sedimentary basins and regions with minimal topographic heterogeneity (Wald and Allen 2007; Allen and Wald 2009; Lemoine et al. 2012). The use of other regional mappable proxies obtained from topographical DEM map, e.g., surface texture and curvature (see Iwahashi and Pike 2007), in conjunction with slope, is likely to allow identifying different landforms, for instance, distinguishing the soil (e.g., depositional sediments) and rock (volcanic lava rock) site conditions despite similar slopes.

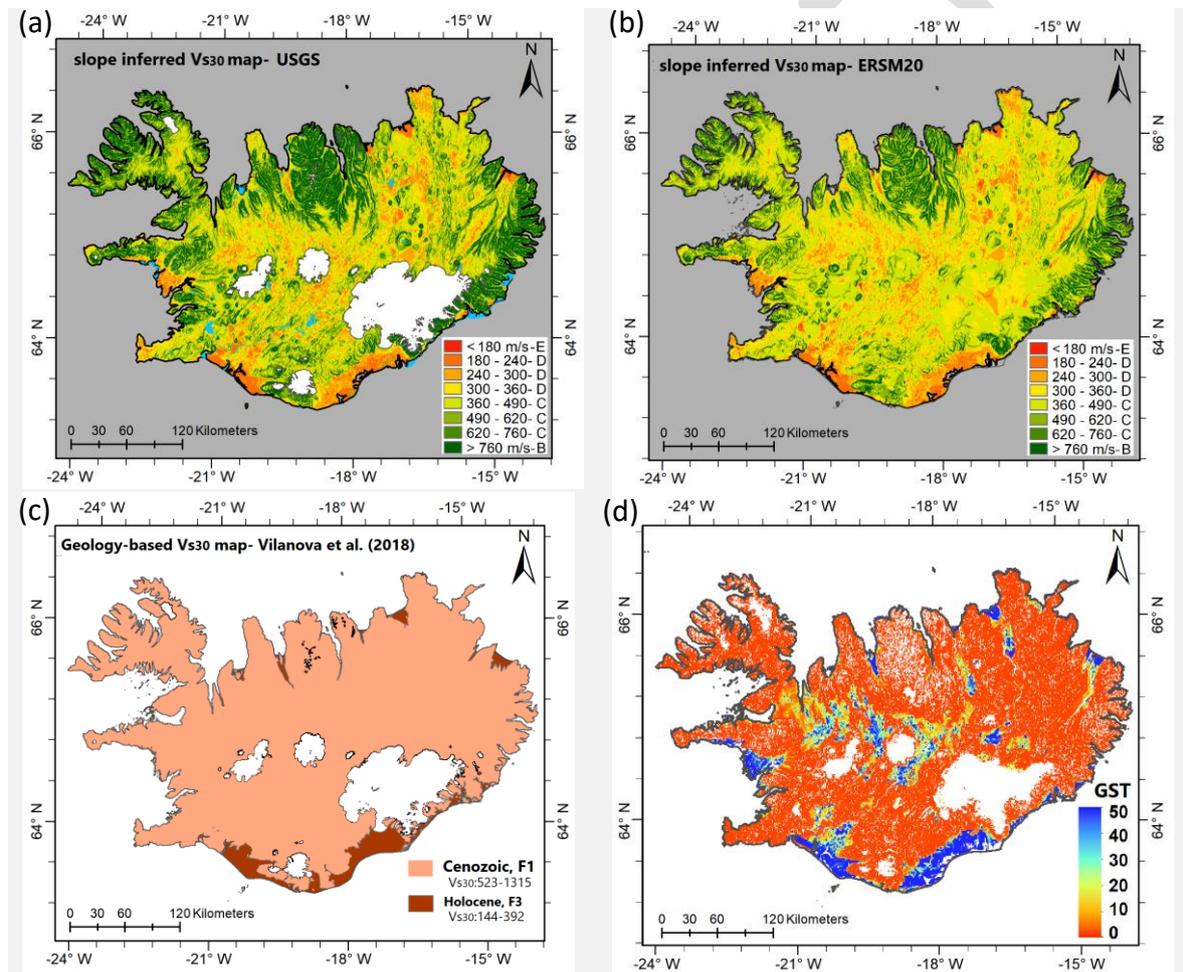

*Figure 4. Site proxy maps for Iceland from large-scale studies: (a) USGS slope-based $V_{S30}$ map developed by Wald and Allen (2007), along with Iceland's water bodies, (b) Inferred ranges of $V_{S30}$ inferred from slope of GEBCO_2014 DEM data, developed for ESRM20´s site response model, (c) geological-based $V_{S30}$ map proposed by Vilanova et al. (2018). For each class, the $V_{S30}$ range is presented by mean ± 1 standard deviation in the legend. (d) Geomorphological Sedimentary Thickness (GST) model developed by Pelletier et al. (2016). (For interpretation of the references to colour in this figure legend, the reader is referred to the online version of this article.)*

### 2.4.2. Geology-based $V_{S30}$ map

The site response model of ESRM20 (Weatherill et al., 2023) employs the stratigraphic classification for Portugal used to build a geology-based $V_{S30}$ model in a passive region (Vilanova et al. 2018) and extrapolated it to the rest of Europe (see Figure 4c). For this, they build a harmonized surface geology map for Europe using three existing geological maps of the European ProMine map at 1:1,500,000, OneGeologyEurope map at 1:1,000,000, and Iceland´s bedrock geological Weatherill et al. (2023). In the end, seven geological eras were proposed, namely: Precambrian, Paleozoic, Jurassic-Triassic, Cretaceous, Cenozoic, Pleistocene, Holocene for every cell of 30-arcseconds resolution in Europe including Iceland and assigned the inferred $V_{S30}$ ranges of Vilanova et al. (2018) to each geological era (Figure 4c).

Figure 4c shows that the majority of Iceland is situated within the Cenozoic era with a smaller portion in the southeastern region falling under Holocene geology. The legend in the figure provides inferred $V_{S30}$ ranges for each geological unit, albeit in broad terms. A comparison between this representation and the geological map of Iceland in Figure 2 highlights the oversimplified geological categorization prevalent in large-scale studies that is overly generalized for conducting local site effect studies. Furthermore, comparing the wide range of inferred $V_{S30}$ values assigned to the geological eras with the only evidence-based $V_{S30}$ values associated with Icelandic geological units shows that such broad and uncertain classification is notably inadequate, even for national-scale investigations.

### 2.4.3. Sediment thickness map

The thickness of soil and sediments down to bedrock is an important factor for modelling amplification of earthquake ground shaking. Pelletier et al. (2016) developed a global geomorphological model with 30 arcsec resolution for average soil and sediment thickness. This model was originally developed for hydrology and ecosystem models and relies on multiple site characterization parameters like topographic slope, lithology, stratigraphy and water table data contributing to seismic amplification effects. The average soil and sediment thickness estimates for upland valley bottom and lowlands were used as a proxy for basin depth in Japan, but no robust trend was found in site model's residuals, the same was observed for the whole Europe (Weatherill et al. (2020), Weatherill et al. (2023)). It was also included in the open-source site database of strong motion stations in Japan (Zhu et al. (2021)). Most recently, it was introduced as a new site proxy called geomorphological sedimentary thickness (GST) to develop a site amplification model for Europe (Loviknes et al. (2024). Figure 4d shows the GST map across Iceland with values ranging from 0 to 50. Comparing this with the spatial distribution of geological structures in Iceland suggests that the most realistic comparison is only observed along the southeast coast, whereas considerable discrepancies are notable in other coastal regions, such as the southwest and west coast, where the surface geology of large regions is in fact Holocene lava rock.

Overall, none of the site characterization proxies introduced for large-scale regions resemble the geological structure of Iceland.

## 3. Site amplification models for Iceland

### *3.1. Quantification of site amplification based on local studies*

Recently, Rahpeyma et al. (2023) analysed all available mainshock strong-motion data in the SISZ from moderate-to-strong earthquakes since 1987. Using one of the new empirical ground motion models (Kowsari et al. 2020) as the backbone model in Bayesian hierarchical modelling they derived station term ($\delta S2S$) estimates over a wide frequency range for 34 strong-motion stations in the SISZ (purple and red triangles in Figure 1d). They showed that the station term residuals reveal a frequency-dependent site amplification at the recording sites. By inspecting and classifying this systematic behaviour, they proposed four $\delta S2S$-groups and grouped the corresponding stations accordingly. Comparison of station locations in each group with the high-resolution geological map (Figure 2b) showed that the geological units and station term groups are correlated. By combining the two, they presented the first quantitative estimates of frequency-dependent average site amplification at four key geological units in Iceland based on age and formation. The four groups are referred to as hard rock, rock, (Holocene) lava rock, and soil (loose deposits). The hard rock (HR) group consists of the oldest basaltic rock (> 3.3 my) and older basaltic and intermediate extrusive rocks and sediments formations from the lower Pleistocene period (0.8-2.5 Ma) (green and dark blue layers in Figure 2, green in Figure 5). Rock group (R) consists of old basaltic lavas with age of <0.8 my (gray layer in Figure 2, blue in Figure 5) and surficial hyaloclastite/rhyolite formation (brown and yellow layers in Figure 2, blue in Figure 5). The lava rock group (L) consists of Holocene lavas (pink and purple layers in Figure 2, purple in Figure 5). The sedimentary soil group (S) consists of looser and younger Holocene sediments (light blue layer in Figure 2, dark gray in Figure 5). As is apparent from Figure 2, these key geological layers are prevalent over the entire country. Using the above grouping, we expand the geology-based site grouping of amplification units in Figure 5 over the entire country. Specifically, each site class is designated to a specific polygon identified based on the available information on the spatial distribution of the detailed geological units. The water bodies like rivers, lakes, oceans and glaciers are shown in white.

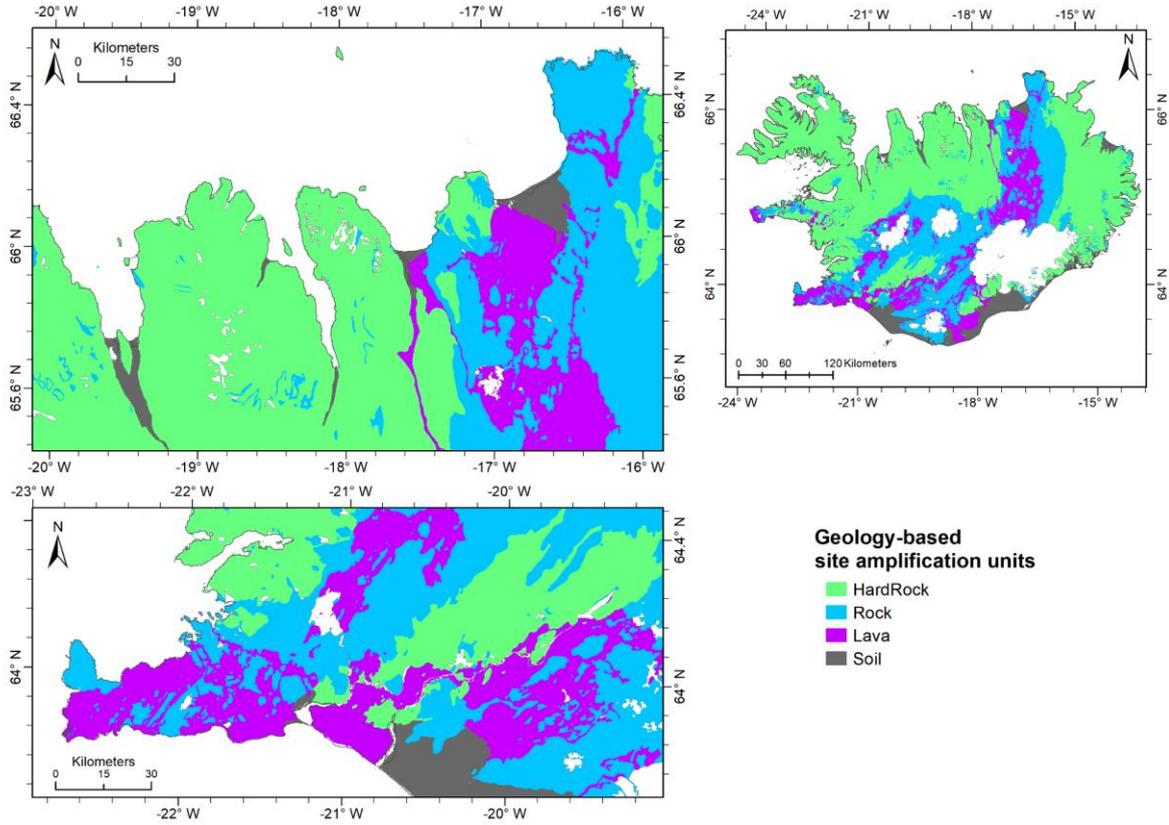

*Figure 5. The proposed site amplification units for the whole Iceland based on four distinct geological-based site classes, namely: hard rock, rock, lava rock and sedimentary soil. (For interpretation of the references to colour in this figure legend, the reader is referred to the online version of this article.)*

### *3.2. Frequency-dependent site amplification models for Iceland for key geological units*

The HR group is considered as a reference site condition since they revealed the lowest site amplification functions. Having the average frequency-dependent site amplification curves for each of the four distinct site classes relative to the reference site condition (Rahpeyma et al. 2023, Fig. 10), along with the detailed geological map of Iceland, we present frequency-dependent site amplification maps at 1, 2, 5, 7, 10-30 Hz along with PGA for the SISZ-RPOR in Figure 6, the TFZ in Figure 7 and the entire country of Iceland in Figure 8.

Each site class polygon is shown in Figure 5 is associated with a posterior distribution of relative frequency-dependent site amplifications (>1.0) or site de-amplifications (<1.0), denoted as $\delta S2S_C^r$ with $C$ representing hard rock, rock, lava rock and sedimentary soil classes. The $\delta S2S_C^r$ is calculated based on the distribution of average station term posteriors of each of the $\delta S2S$-groups at different frequencies relative to the average reference station term (i.e., average of median posterior estimates of $\delta S2S$ for reference site stations). For the sake of clear illustration, the frequency-dependent site amplification maps are color-coded by a specific range of $\overline{\delta S2S_C^r}$ (median estimate of $\delta S2S_C^r$ posterior distribution at each C site class) as shown in the legend of Figure 6 to Figure 8.

As shown in Figure 8a, the amplification function associated with four geological-based site classes do not differ at frequency of 1 Hz, as the $\overline{\delta S2S_C^r}$ for all site classes are almost equal to 1.0, thus, only one-color amplification map is displayed. This is also the case for all frequencies lower than 1 Hz (see $\overline{\delta S2S_C^r}$ for 1, 0.5 and 0.3 Hz in Figure 11b of Rahpeyma et al. (2023)). However, for frequencies larger than 1 Hz, the amplification factors associated with different geological units vary systematically and considerably with broader dispersion indicating the actual variation of station term classes.

Figure 8b illustrates the site amplification map at 2 Hz in two colors, indicating two distinct groups of reference and rock sites vs. lava and soil sites, although their corresponding posterior distributions overlap. At frequencies larger than 3 Hz, the relative site amplification value of soil class increases at a higher rate and separates from the lava rock class and then above 5 Hz, therefore, creating three distinct groups as shown by the site amplification map at 5 Hz (i.e., hard rock and rock site amplifications are identical). We should note that the posterior distributions of the four site classes overlap to some extent. At frequencies above 5 Hz, the hard rock site class segregates from rock site and effectively four distinct site classes appear. This can be seen in site amplification maps at 7 Hz and also at 10-30 Hz where the four geological-based site classes are evidently distinguishable. We note that at 10 Hz to 30 Hz, the amplification factors relative to the reference site conditions of the four classes are almost constant with the approximate values of 3.0, 2.0, and 1.5 for soil, lava rock, and rock site classes, respectively. It should be highlighted that the hard rock (reference) site class (green polygon layer in Figure 5) is always shown by a dark green color in Figure 6 to Figure 8, due to the fact that the reference site class average amplification is near unity across the whole frequency range of interest. At PGA, the differences of relative site amplification values of the lava rock and rock classes decrease, resulting in three amplification levels.

The frequency-dependent site amplification models presented in this study enable further investigation on spatial variability of ground motions predictions across the whole country accounting for geological-based site class amplifications along with their uncertainties.

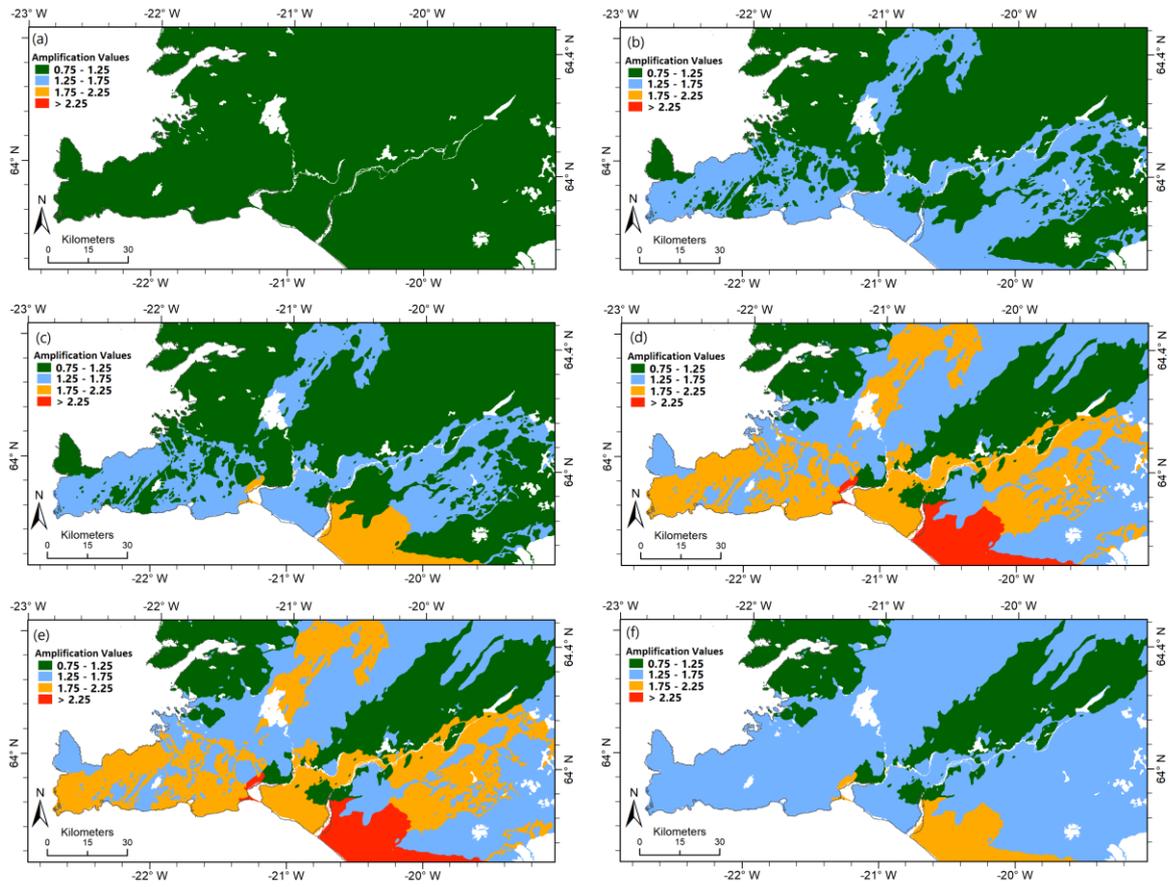

*Figure 6. Frequency-dependent site amplification models at 1 (a), 2 (b), 5 (c), 7 (d), 10-30 Hz (e) and PGA (f) for SISZ-RPOR seismic fracture zone. The amplification maps are relative to the median prediction of Rahpeyma et al. (2023). (For interpretation of the references to colour in this figure legend, the reader is referred to the online version of this article.)*

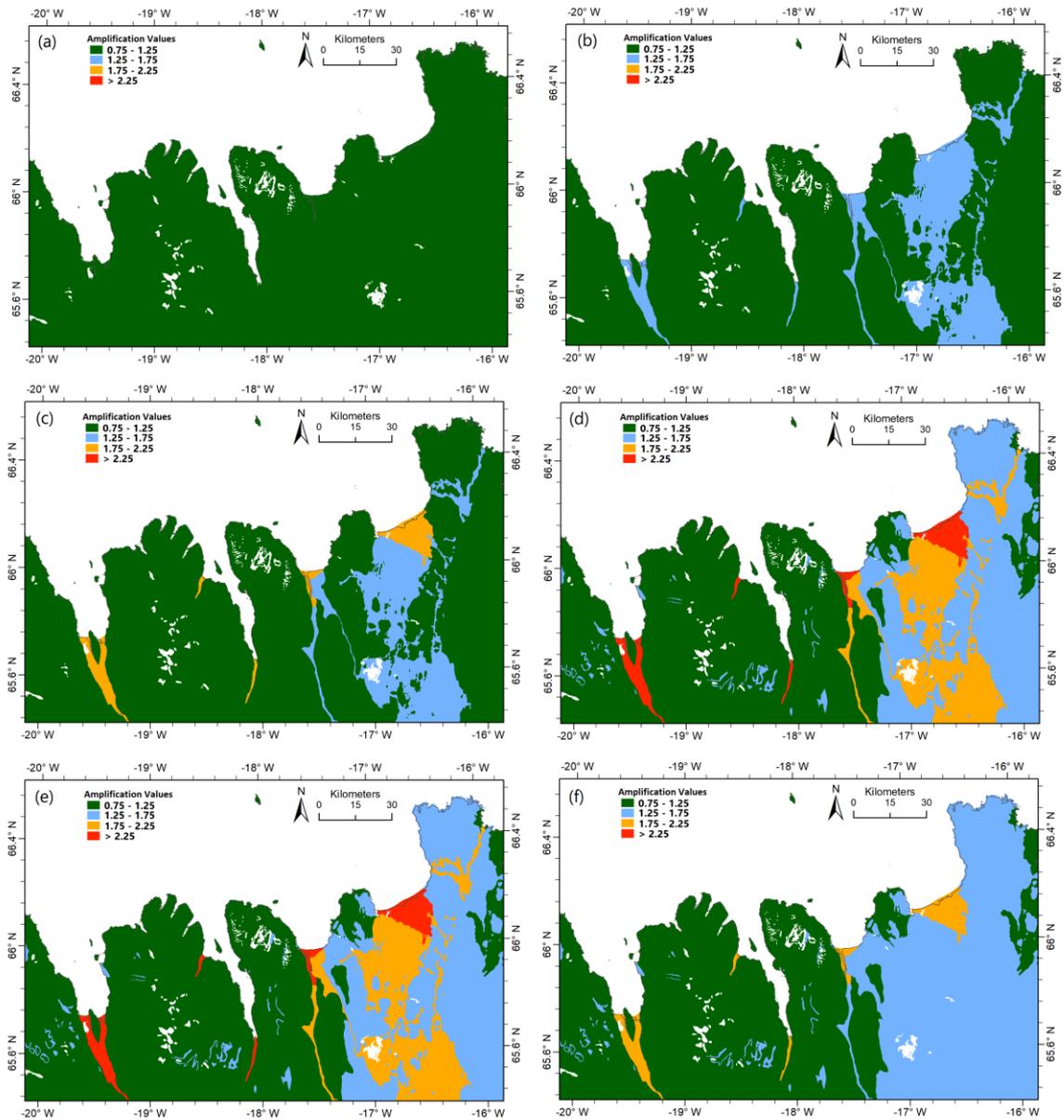

*Figure 7. Frequency-dependent site amplification models at 1 (a), 2 (b), 5 (c), 7 (d), 10-30 Hz (e) and PGA (f) for TFZ seismic fracture zone. The amplification maps are relative to the median prediction of Rahpeyma et al. (2023). (For interpretation of the references to colour in this figure legend, the reader is referred to the online version of this article.)*

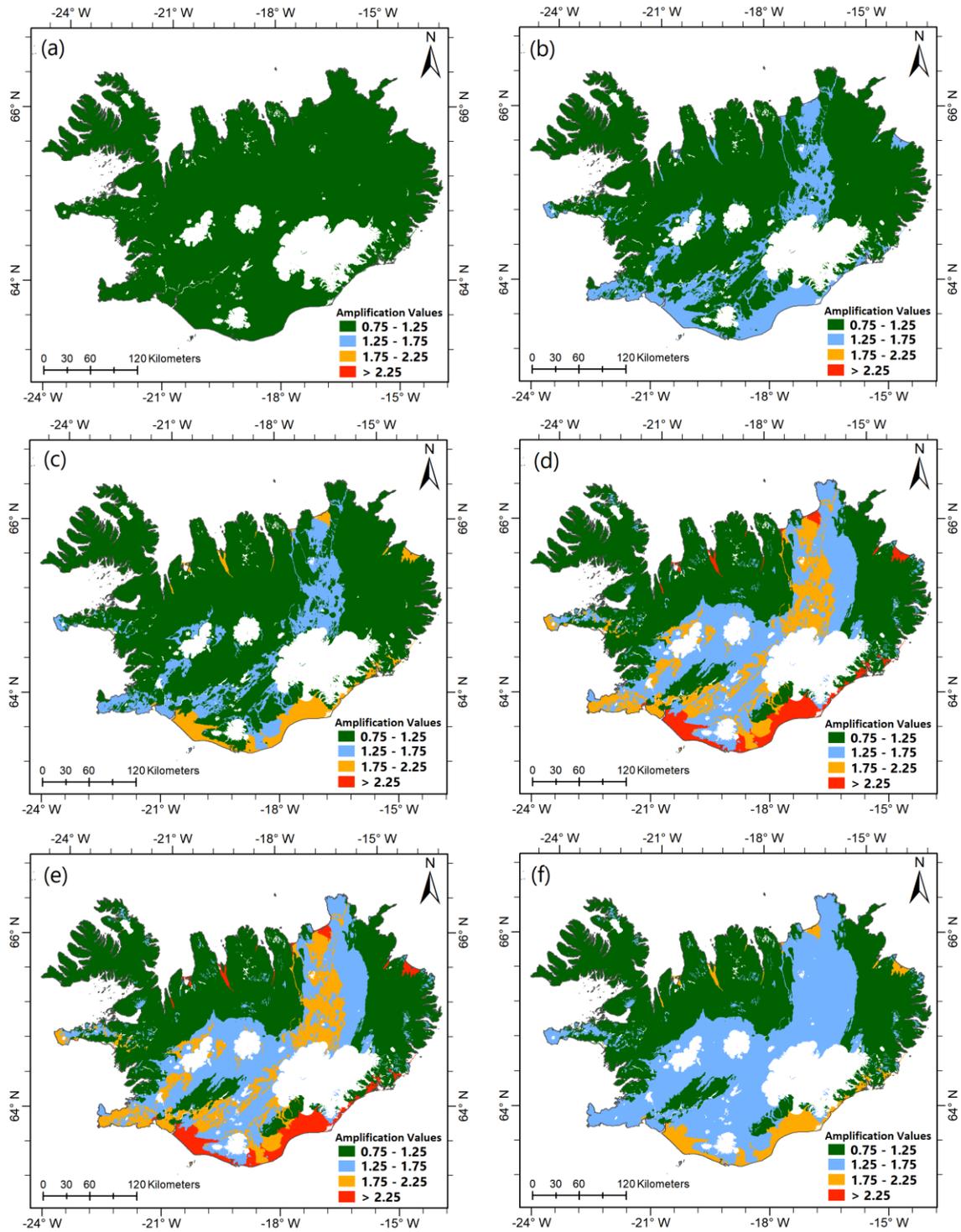

*Figure 8. Frequency-dependent site amplification models at 1 (a), 2 (b), 5 (c), 7 (d), 10-30 Hz (e) and PGA (f) for Iceland. The amplification maps are relative to the median prediction of Rahpeyma et al. (2023). (For interpretation of the references to colour in this figure legend, the reader is referred to the online version of this article.)*

### 3.3. Spatial pattern of local estimates of site amplification vs. $V_{S30}$ estimates

As indicated in Table 1, the Geological-based $V_{S30}$ map of Iceland (Figure 3) combines the old basaltic bedrock with age of 0.8-3.3 my and greater than 3.3 my (dark blue and green layers in Figure 2) into a single $V_{S30}$ class of ~2000 m/s (green layer in Figure 3). We also apply a similar approach to amplification classes, as Hard Rock is represented by green layer in Figure 5.

The $V_{S30}$ map distinguishes between Holocene lavas and those in the South Iceland Lowland, presumed to have flowed over sedimentary plains (~1500 vs ~1300 m/s). However, no distinction is made between them based on our $\delta S2S$ data (Lava group in Figure 5).

The $V_{S30}$ map unifies hyaloclastite/rhyolite formations older than 11,000 years (younger) with the older ones (yellow and brown layers in Figure 2) into a single $V_{S30}$ class of ~1200 m/s. The same happens concerning the site amplification class of Rock (blue layer in Figure 5), which also encompasses the <0.8 my basaltic lavas with $V_{S30}$ of ~1600 m/s, albeit with limited data on these formations. It's worth noting that the former unit is highly localized in the highlands. The $V_{S30}$ map distinguishes between the deep and extensive Holocene sedimentary formations along the south coast (~600 m/s) and shallower, localized soil deposits in the north (~700 m/s). However, both are considered soil groups in the site amplification map in Figure 5 (gray).

Table 1: *The frequency-dependent $\delta S2S$ relative amplification factors associated with Iceland's geological units along with their corresponding local estimates of $V_{S30}$. The background color of the amplification factors is consistent with their corresponding polygon in the frequency-dependent site amplification maps. The amplification factors are relative to the median prediction of Rahpeyma et al. (2023).*

| Site class | Frequency | | | | | | $V_{S30}$ (m/s) | Geological unit |
| --- | --- | --- | --- | --- | --- | --- | --- | --- |
| | 1 Hz | 2 Hz | 5 Hz | 7 Hz | 10-30 Hz | PGA | | |
| Lava | 0.75-1.25 | 1.25-1.75 | 1.25-1.75 | 1.75-2.25 | 1.75-2.25 | 1.25-1.75 | ~1500 | Lavas (basaltic, <1.1ky) |
| Lava | 0.75-1.25 | 1.25-1.75 | 1.25-1.75 | 1.75-2.25 | 1.75-2.25 | 1.25-1.75 | ~1500 | Holocene Lavas (basaltic, >1.1 ky) |
| Lava | 0.75-1.25 | 1.25-1.75 | 1.25-1.75 | 1.75-2.25 | 1.75-2.25 | 1.25-1.75 | ~1300 | Holocene Lava- <25m top sediment thickness |
| Rock | 0.75-1.25 | 0.75-1.25 | 0.75-1.25 | 1.25-1.75 | 1.25-1.75 | 1.25-1.75 | ~1200 | Hyaloclastite, rhyolite (>11 ky) |
| Rock | 0.75-1.25 | 0.75-1.25 | 0.75-1.25 | 1.25-1.75 | 1.25-1.75 | 1.25-1.75 | ~1200 | Hyaloclastite, rhyolite (<0.8my) |
| Rock | 0.75-1.25 | 0.75-1.25 | 0.75-1.25 | 1.25-1.75 | 1.25-1.75 | 1.25-1.75 | ~1600 | Basaltic lava (<0.8 my) |
| HR | 0.75-1.25 | 0.75-1.25 | 0.75-1.25 | 0.75-1.25 | 0.75-1.25 | 0.75-1.25 | ~2000 | Basaltic rock (0.8-3.3 my) |
| HR | 0.75-1.25 | 0.75-1.25 | 0.75-1.25 | 0.75-1.25 | 0.75-1.25 | 0.75-1.25 | ~2000 | Basaltic rock (>3.3 my) |
| Soil | 0.75-1.25 | 0.75-1.25 | 1.75-2.25 | > 2.25 | > 2.25 | 1.75-2.25 | ~600 | Holocene Sediments, mostly alluvium, thickness > 30m |
| Soil | 0.75-1.25 | 0.75-1.25 | 1.75-2.25 | > 2.25 | > 2.25 | 1.75-2.25 | ~700 | Holocene Sediments, thickness < 30m |

# 4. Comparison with proxy-based site amplification maps from large-scale (non-Icelandic specific) studies

## *4.1. European site amplification models*

Weatherill et al. (2023) developed regional site amplification models for shallow seismicity in Europe for the purpose of ESRM20 advanced by accounting for geological era in combination with a site predictor variable that can be either measured $V_{S30}$, inferred $V_{S30}$, or topographic slope itself. Most recently, Loviknes et al. (2024) derived four sets of site amplification models for Europe and Türkiye based on four well-known site proxies of topographically inferred $V_{S30}$ of USGS (Figure 4a), slope itself, combining geological era (Figure 4c) and slope, and finally GST (Figure 4d). Unlike ESRM20, they used Fourier Amplitude Spectra to derive simple GMMs following the procedure of Kotha et al. (2022), in order to more effectively capture the physical effects that might be obscured in the response spectra, particularly at high frequencies (Kotha et al. 2022). The site amplification models are derived from empirical site-to-site ground motion residuals ($\delta S2S_s$) which represent the systematic deviation of recorded ground motions from the GMM median predictions at a specific site calibrated to the European Engineering Strong-Motion (ESM) dataset (Lanzano et al. 2017; Luzi et al. 2020). Unlike common site-amplification factors like the factors shown in Figure 6 to Figure 8, their $\delta S2S_s$ is not relative to a reference site condition, but to $\delta S2S_s$ = 0 (i.e., center of the distribution) that is the median of all the stations.

Figure 9 shows the mean site amplification maps at 2 Hz derived from the four above-mentioned proxies. We note that the amplification predictions should be interpreted in reference to the median prediction of the associated GMM and considering that the site-to-site term is relative to the median of all stations across Europe and Türkiye.

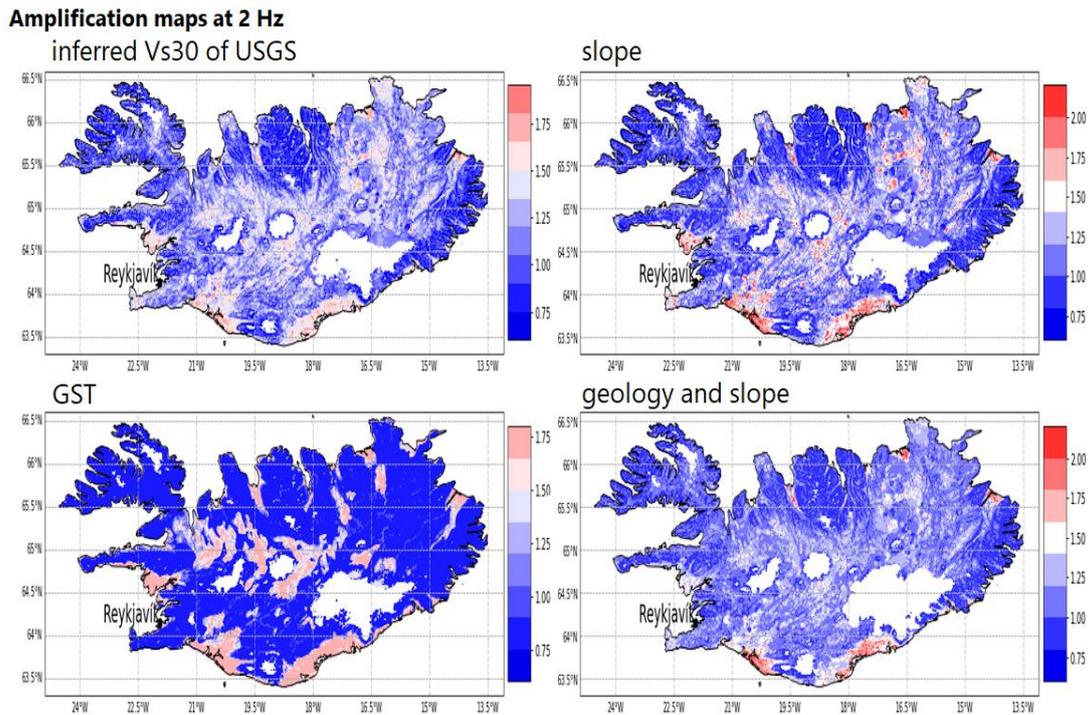

*Figure 9. Proxy-based site amplification maps at 2 Hz proposed by Loviknes et al. (2024). Each panel includes predictions based on four different proxies as shown in the subplot's title, namely: slope-based inferred $V_{S30}$ of USGS, slope gradient, Geomorphological Sedimentary Thickness (GST), and combination of geological era and slope gradient. The amplification maps are relative to the median prediction of Loviknes et al. (2024). (For interpretation of the references to colour in this figure legend, the reader is referred to the online version of this article.)*

European proxy-based site amplification models peak at low frequencies ($f$ < 1 Hz). In other words, none of them shows distinguishable differences at high frequencies ($f$ > 3 Hz) which is most likely attributed to the low resolution of proxies (e.g., poor geological map) (Loviknes et al. 2024; Weatherill et al. 2023). Recent site response investigations for Japan and Switzerland indicate that topographic, geological site proxies exhibit a strong correlation with low-frequency site response (0.5-3.3 Hz). Particularly, larger-scale (1140-2020 m) topographical parameters prove most effective for lower frequency amplification factors (0.5-1 Hz) (Bergamo et al. 2021). This is in stark contrast with our quantified site amplification models exhibiting pronounced amplification peaks at high frequencies of above 7-10 Hz which strongly rely on the geological attributes (see Figure 6 to Figure 8), also confirmed by resonance at short periods (T < 0.5 s) observed on the inter-station posterior distribution of most stations in SISZ. Additionally, as low-frequency motions damp drastically due to Icelandic geological features, the geology-based site classes (HR, R, L, S) show insignificant differences at frequencies less than 1 Hz which is in great disagreement with large-scale non-Icelandic site amplification models. Care should be taken to interpret large-scale models with relatively low-resolution data with limited observations on specific geological eras. For instance, in ESRM20, stations associated with older geological eras are mainly located in mountainous areas in Europe while stations on Holocene sediments relates to flat local topography with higher slopes tend to be located at the edges of valleys (mostly in upland areas) (Weatherill et al. 2023). We also highlight that as the European site amplification maps are associated with median predictions of their respective GMMs, for the purpose of fair comparison, they are compared qualitatively with our local models. Nonetheless, their significant disagreement in terms of the spatial and spectral patterns of amplification factors makes any additional calibration of GMMs unnecessary.

### *4.2. Exploring site-to-site variability in SISZ*

Having the empirical original $\delta S2S_s(f)$ at 34 strong motion stations ($s$) in SISZ, we evaluate the prediction capability of the large-scale non-Icelandic site amplification models and compare them with the Icelandic one, i.e., Rahpeyma et al. (2023), hereinafter Rea23 (Figure 10). The large-scale models are four proxy-based models from Loviknes et al. (Lea24, 2024) (see Figure 9) and an ESRM20 (Weatherill et al., 2023) model based on slope and geology.

For this, we compute the residual (corrected) site response ($\delta S2S_{s,cor}^m$, in log-10 scales) between observations ($\delta S2S_s^0(f)$) and a proxy-based prediction model, $m$ ($\delta S2S_s^m(f, proxy)$) at frequencies ranging from 1 to 10 Hz as follows:

$$\delta S2S_{s,cor}^m(f, proxy) = \delta S2S_s^0(f) - \delta S2S_s^m(f, proxy) \qquad (1)$$

$$\delta S2S_{s,cor}^m(f, proxy) \sim N(0, \varphi_{S2S,cor}^m(f, proxy)) \qquad (2)$$

Assuming a normal distribution for $\delta S2S_{s,cor}^m$ with zero mean, $\varphi_{S2S,cor}^m$, the (corrected) residual site-to-site variability, is its corresponding standard deviation. $\delta S2S_{s,cor}^m$ represents part of the $\delta S2S_s^0(f)$ that is not modelled by the proxy-based prediction model, $m$ (Weatherill et al. 2020b). Therefore, $\varphi_{S2S,cor}^m$ indicates the site-to-site variability in the residual site response not captured by model $m$ at 34 stations in SISZ. An ideal prediction model would result in almost zero $\varphi_{S2S,cor}^m$. Therefore, comparing the reduction in the $\varphi_{S2S,cor}^m$ can be an indicator of the efficiency of the $m$ proxy-based model in predicting the empirical site amplification (Stewart et al. 2017; Zhu et al. 2022).

The results presented in Figure 10 follow the same presentation to that of Lea24, but here we present two measures of model performance. First, the $\text{bias}_m(f)$ of each model as a function of frequency is shown as solid lines, representing the absolute value of the mean of the model residuals at each frequency, i.e., $\text{bias}_m(f) = |\text{mean}(\delta S2S_{s,cor}^m(f))|$. In other words, it is a measure of how much each model under/overpredicts the observations on average. The Rea23 model is the only one that can be viewed as being unbiased across the frequency range i.e., the mean of its model residuals is near zero (ranging from 0.0015 to 0.009 over all frequencies, they are insignificantly different from zero). That contrasts with most of the other models that show a large bias against the observations. Two exceptions are the Lea24-slope and Lea24-inferred $V_{S30}$ models, both of which have zero bias, but only at $f > 2$ Hz. At lower frequencies, however, these models are biased against the observations.

Second, the dashed lines represent the bias of each model plus its standard deviation, i.e., $\text{bias}_m(f) + \varphi_{S2S,cor}^m(f)$. In other words, the dashed lines give an idea of how far from the zero line each model bias and its residual variation is. Again, the Rea23 model with $\varphi_{S2S,cor}^{\text{Rea23}}$ of 0.05-0.122 shows the least scatter i.e., average differences from the observations. In contrast, we name two examples. The ESRM20 model has a very small standard deviation at 1 Hz (~0.05), but a very large bias (~0.21). However, the two Lea24 models mentioned above have effectively zero bias at 7 Hz, but very large standard deviations (~0.20).

The results therefore show that in addition to Rea23 being unbiased over the frequency range, it also fits the data better i.e., has the smallest standard deviation of all the models.

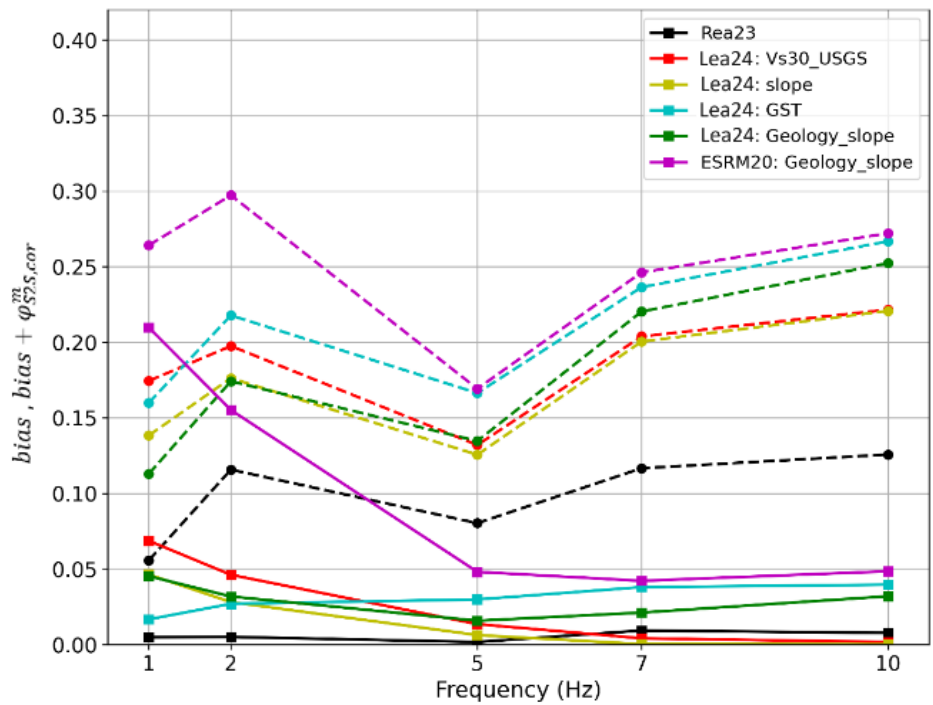

*Figure 10. The corrected (residual) site-to-site variability, $\varphi_{S2S,cor}^m$ corresponding to Icelandic site amplification model from Rahpeyma et al. (Rea23, black curves), four proxy-based site amplification models of Loviknes et al. (Lea24) and an ESRM20 model (with reference slope of 0.3 m/m). The solid lines indicate bias, i.e., absolute value of mean $\delta S2S_{s,cor}^m$ at $f$. The dashed lines show $bias + \varphi_{S2S,cor}^m$. (For interpretation of the references to colour in this figure legend, the reader is referred to the online version of this article.)*

Given the inherent uncertainty associated with both proxy-based amplification prediction models and observed δS2S, Figure 11 compares the amplification predictions with observed station terms and their variability at nine strong motion stations in the SISZ. The selection of stations was based on their representation of different geological units, including Hard Rock, Rock, Lava, and Soil. In addition, they recorded data for the six earthquakes ($Mw$5.1-6.5) that were used to calculate the originally observed δS2S posteriors. The error bars for the four proxy-based amplification factors from Lea24 denote the mean ± 1 standard deviation of model coefficients, and the black error bar of the local geology-based model of Rea23 represents the associated 16$^{th}$ - 84$^{th}$ percentile range. The red error bar denotes the 16$^{th}$ - 84$^{th}$ percentiles of the posterior of the observed δS2S obtained from the recordings of 6 earthquakes in SISZ. Notably, the Lea24 model based on the combination of geology and slope has the highest uncertainty while the models based on GST and slope (cyan and yellow error bars) have the lowest uncertainty range. Consistent with the results for site-to-site variability, there is considerable agreement between the Rea23 prediction ranges and the observed ranges of site factors at all stations shown in *Figure 11*. The largest discrepancy is observed for the prediction models based on GST and inferred $V_{S30}$. This comparison helps us to understand how reliable proxy-based amplification models are in capturing the actual site factors at given stations.

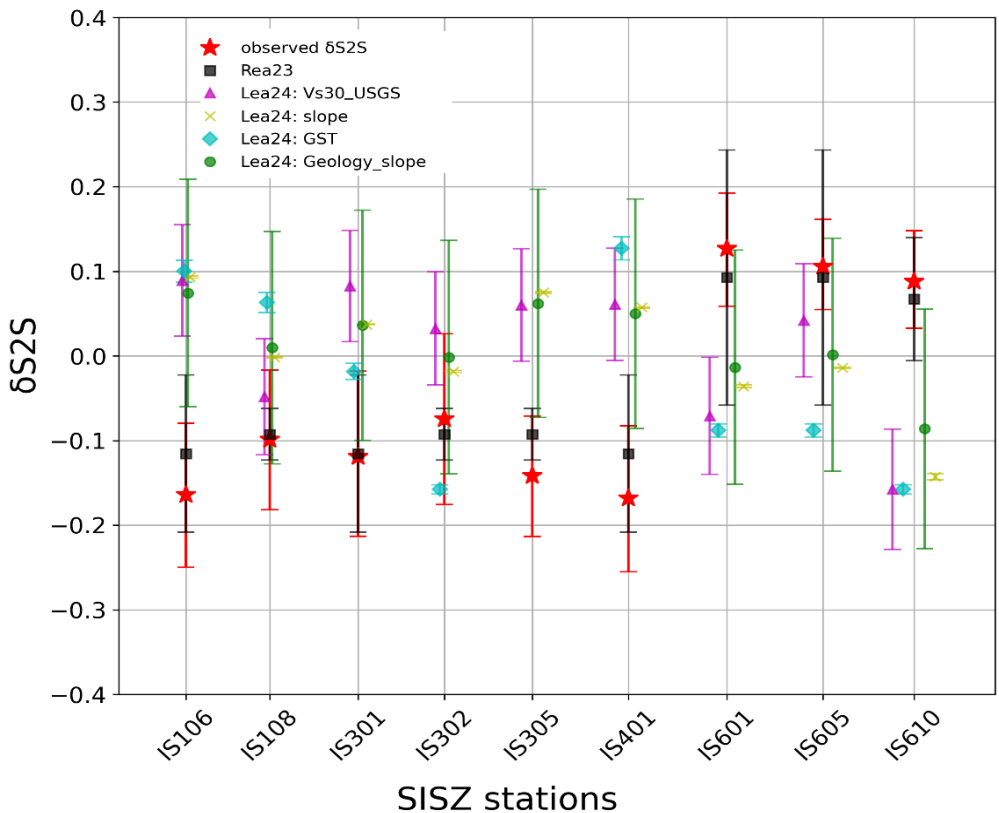

*Figure 11. Comparison of the proxy-based amplification predictions with observed station term at nine SISZ strong motion stations*

## 5. Conclusions

This study builds on the first rigorous quantification of frequency-dependent site amplification factors correlated to an independent predictor i.e., distinct Icelandic geological units, that resulted in the first empirical ground motion model for Iceland that accounts for seismic wave amplification other than the qualitative site specification of "rock" and "stiff soil" (Rahpeyma et al. 2023). The models developed based on strong-motion stations in Southwest Iceland using 83 mainshock ground motions recordings ($M_w$5.1-6.5, $R_{JB}$=1-100 km). Specifically, the geological units were specified as hard rock, rock, lava rock, and sedimentary soil; and frequency-dependent site-amplification functions were presented relative to the hard rock condition. However, while the abovementioned study applied to Southwest Iceland, the geological units are prevalent throughout the country. In this study, therefore, we developed maps of these four site classes for entire Iceland for which the site-specific empirical ground motion model applies, with particular focus on the two transform zones of the country, on the basis of digital elevation models and detailed geological maps. We present maps of site amplification for the entire country, relative to the very hard rock site condition and at various frequencies, including 1, 2, 5, 7, 10-30 Hz, and PGA.

In addition, we have produced nationwide maps of common site-effect proxies such as geological units, geological age, topography and slope. For completeness, and due to the worldwide popularity of the $V_{S30}$ measure as a proxy for site effects, we revive and regenerate a $V_{S30}$ map of Iceland developed based on a collection of shallow shear wave velocity estimates and associated with key geological units in Iceland. The maps of quantitative site amplification and the original $V_{S30}$ maps now allow the association of the site effect proxy to the amplification.

For comparison, we produce multiple nationwide maps of predicted site amplification for Iceland based on models developed in other seismic regions and/or applied in other studies (e.g., ESRM20, Loviknes et al. (2024)) along with various site proxies they are based on (e.g., geological- and slope-based inferred $V_{S30}$, including the most recent one, geomorphological sedimentary thickness). In terms of comparing the site proxies, as expected, there is a significant difference between local models presented herein and large-scale proxies such as detailed high-resolution geological maps and local $V_{S30}$ estimates associated with the detailed geological features. In addition, we qualitatively compare the European proxy-based amplification maps with our nationwide site amplification models developed based on local estimates of site-to-site strong ground motion residuals. The European amplification factors are derived in relation to their corresponding GMM median predictions, necessitating additional calibration for a rigorous quantitative comparison. However, it is imperative to underscore that, even from a qualitative standpoint, stark disparity exists between local and both large-scale models and therefore providing a detailed account of the comparison is unnecessary. It suffices to state that neither spatial patterns nor spectral trends of amplification factors from large-scale non-Icelandic studies resemble those observed in this study. We compared model performance across frequencies by assessing the bias of model predictions against Icelandic empirical site amplification estimates in SISZ, accounting for site-to-site variability of residuals. The results demonstrate a superior performance of the Icelandic amplification model in capturing the observed data.

Overall, we attribute the observed discrepancy primarily to the young geology of Iceland and its formations, differing significantly from the regions used in the data collection for the large-scale studies. That in turn emphasizes the importance of considering local geological intricacies in site effect studies. Addressing these challenges is imperative for enhancing the reliability and applicability of seismic hazard assessments in Iceland and beyond. This study, meanwhile, provides comprehensive seismic site amplification maps for Iceland based on quantitative site-effect studies using the available dataset of Icelandic mainshock strong-motions. Moreover, it revives for the sake of comparison with other regions worldwide the expected spatial pattern of $V_{S30}$ values in a young volcanic region of Iceland. The two allow a more informed estimation of earthquake ground motion amplitudes that is expected to provide more optimal and reliable estimates of seismic hazard in Iceland.

## Data and resources

The raster datasets of the proposed frequency-dependent site amplification maps for Iceland are provided on GitHub (https://github.com/AtefeD/Iceland_SiteAmpMap). Full detail characteristics of the Icelandic recording stations in SISZ including geological units, site amplification factors, the $\delta S2S_s$ statistic along with all associated site effect proxies at each station are given as well. A Python code for computing and mapping the proxies and site-amplification prediction maps is provided on GitHub (https://github.com/AtefeD/Iceland_Site_Characteristics).

Detailed geological map of Iceland can be viewed through ÍSOR webmap service here https://arcgisserver.isor.is/?lon=-19.00213&lat=65.00348&zoom=7&layers%5B%5D=satellite. The proxy-based site amplification models proposed by Loviknes et al. (2024) are available at: https://doi.org/10.5281/zenodo.8072116 (last accessed 20.05.2023). The 30-arcsecond gridded thickness of geomorphological sedimentary deposit layer estimated by Pelletier et al. (2016) can be downloaded from https://daac.ornl.gov/cgi-bin/dsviewer.pl?ds_id=1304 (last accessed 20.05.2023). The slope-based inferred $V_{S30}$ raster file developed by Wald and Allen (2007) can be downloaded from the U.S. Geological Survey (USGS) via https://earthquake.usgs.gov/static/lfs/data/vs30/vs30.zip (last accessed 20.05.2023). The slope and geology-based site amplification map used in the European Seismic Risk Model (ESRM20) are constructed using the open source 'Exposure To Site' tool available here: https://gitlab.seismo.ethz.ch/efehr/esrm20_sitemodel. The key datasets of slope, geological era, and inferred $V_{S30}$ from slope that were used in ESRM20 are provided through web-service of EFEHR European Site Response Model Datasets Viewer (https://maps.eu-risk.eucentre.it/map/european-site-response-model-datasets/#4/53.98/4.53) as well as from https://nextcloud.gfz-potsdam.de/s/93ZR4ky8D4mDXb9 (last accessed 20.05.2023). All maps presented in this study are generated using ArcGIS 10.8.

## CRediT authorship contribution statement

**Atefe Darzi**: Conceptualization, Methodology, Formal analysis, Investigation, Map generation, Writing - Original Draft, Funding acquisition, Revision. **Benedikt Halldorsson**: Conceptualization, Methodology, Resources, Writing - Review & Editing, Supervision, Project administration, Funding acquisition, Revision. **Fabrice Cotton**: Methodology, Writing - Review & Editing, Revision. **Sahar Rahpeyma**: Resources.

## Declaration of competing interest

The authors declare that they have no known competing financial interests or personal relationships that could have appeared to influence the work reported in this paper.

## Acknowledgments

This work was supported by the Icelandic Centre for Research (Rannís) Research Grant (No. 228782) and Postdoctoral Fellowship Grant (No. 218255). It was also facilitated by an Erasmus+ mobility grant for the lead author to visit the GFZ, German Research Centre for Geosciences, Germany. I am grateful to Karina Loviknes and Graeme Weatherill for their development of the European models that served as benchmarks enabling insightful comparisons in this study, as well as for their assistance during my visit, offering clarifications that enriched my understanding. The study was also partially supported by the European Union's Horizon 2020 Research and Innovation program (G.A. No. 823844, "ChEESE"; No. 821046, "TURNkey") and Horizon 2021 EuroHPC JU-RIA project (G.A. No. 101093038, "ChEESE-2P"), the Landsvirkjun Energy Research Fund (No. NÝR-04-2023), the Icelandic Road and Coastal Administration Research Fund (No. 1800-947), and the University of Iceland Research Fund (No. 92334). Finally, we thank two anonymous reviewers for their valuable comments which contributed to improving the quality of the paper.